\documentclass{article}

\usepackage{makeidx}
\usepackage{amsthm,amsmath,amssymb,amsfonts}
\usepackage{setspace,graphicx}
\usepackage{Generic}  %
\usepackage[sort, longnamesfirst]{natbib}
\usepackage{url} 
\usepackage{listings}
\usepackage{multirow}
\usepackage{DejaVuSansMono}

\usepackage[usenames,dvipsnames,table]{xcolor}
\usepackage{tikz}

\usepackage{changepage}

\usepackage{hyperref}
\hypersetup{
    colorlinks=true,
    linkcolor=red,
    citecolor=MidnightBlue,
}

\definecolor{charlesBlue}{RGB}{100, 155, 255}
\definecolor{lightgray}{RGB}{200, 200, 200}

\numberwithin{equation}{section}
\theoremstyle{plain}

\usepackage[capitalize,noabbrev]{cleveref}

\usepackage{setspace}

\begin{document}
\onehalfspacing

\title{\bf{Running Markov Chain Monte Carlo on Modern Hardware and Software}}
\author{Pavel Sountsov\\
{\small Google DeepMind} \\[0.1in]
Colin Carroll\\
{\small Google DeepMind} \\[0.1in]
Matthew D. Hoffman\\
{\small Google DeepMind}
}
\date{}

\maketitle

\begin{abstract}
Today, cheap numerical hardware offers huge amounts of parallel computing power, much of which is used for the task of fitting neural networks to data.
Adoption of this hardware to accelerate statistical Markov chain Monte Carlo (MCMC) applications has been much slower.
In this chapter, we suggest some patterns for speeding up MCMC workloads using the hardware (e.g., GPUs, TPUs) and software (e.g., PyTorch, JAX) that have driven progress in deep learning over the last fifteen years or so.
We offer some intuitions for why these new systems are so well suited to MCMC, and show some examples (with code) where we use them to achieve dramatic speedups over a CPU-based workflow.
Finally, we discuss some potential pitfalls to watch out for.

\end{abstract}

\section{Introduction}

Markov chain Monte Carlo (MCMC) methods first came to be widely used for statistical applications in the 1990s, following the pioneering work of \citet{Gelfand:1990}.
Over the next couple of decades, the statistics and machine-learning communities developed some standard MCMC workflows, and single-processor computers got faster at an exponential rate by increasing clock rates and cache sizes.
Then, around 2005--2010, computer architecture took a left turn: while Moore's law \citep{Moore:1965} continued to hold for transistor density, power and heat-dissipation requirements forced chip makers to increase throughput by also giving processors more parallel-processing power.
Among other things, this led to the broad adoption of Graphics Processing Units (GPUs, originally solely used to accelerate computer-graphics applications) for general-purpose numerical computing.
A consumer-grade GPU that one might find in a typical mid-level computer configuration can perform tens of trillions of floating-point operations (FLOPs) per second; a typical CPU can do a small fraction of that.
Elsewhere, specialized parallel hardware has been developed to accelerate deep-learning applications;
for example, the Tensor Processing Unit \citep[TPU;][]{Jouppi:2017, Jouppi:2023}, can be even more efficient than a GPU for appropriate workloads. GPUs and TPUs can both be rented for a couple of dollars per chip-hour from cloud-computing providers.
``Parallel accelerators'' such as GPUs and TPUs have changed the landscape of computing---particularly for deep learning---but they are often under-utilized in statistical workloads like  MCMC.
This is although such workloads are a good fit for this new world of specialized parallel accelerators;
MCMC computations often offer opportunities to parallelize computation across data, model parameters, and chains.

Cheap parallel accelerators are only part of the picture; a program's computations may need to be radically reorganized to take advantage of parallel hardware.
Fortunately, there is a vibrant ecosystem of open-source software frameworks (mostly developed to support the deep-learning community) that dramatically simplify the task of writing numerical code that runs efficiently on parallel acceleration. Today, two of the most popular of these libraries are PyTorch \citep{Paszke:2019}, and JAX \citep{JAX:2018}.
JAX conveniently exposes an interface that closely resembles NumPy \citep{Harris:2020}, which is widely used for scientific computing in the Python ecosystem.
For this chapter, we assume some basic knowledge of NumPy and Python and will point out where there are JAX-specific extensions and limitations.
Most of the considerations we discuss apply equally to other Python deep-learning libraries, and other languages as the underlying hardware restrictions of parallel accelerators remain the same.

Writing numerical code in these frameworks is a lot like writing code for more CPU-oriented numerical computing environments like MATLAB \citep{Matlab}, R \citep{R}, and NumPy.
There is a library of numerical functions that accept and return multidimensional arrays (e.g., a vector is a one-dimensional array, a matrix is a two-dimensional array, etc.).
The dimensions are often referred to as ``axes" (e.g. a 2D matrix with $n$ rows and $d$ columns representing a dataset of features could be said to have an ``examples'' axis with size $n$ and a ``feature'' axis with size $d$).
There are vectorized operations which operate over one or more axes: for example, a mean function can compute the mean of the entire 2D matrix, or the per-feature or per-example mean by operating over the feature or example axis, respectively.

The framework is responsible for figuring out the low-level implementation details of these vectorized computations, and in particular for ensuring that they are run efficiently on whatever specialized hardware (e.g., a GPU) is available.
These frameworks also offer facilities for automatic \emph{program transformations}, notably automatic-differentiation and vectorized-map operations; the former makes it much easier to use gradient-based MCMC methods based on Hamiltonian Monte Carlo \citep[HMC;][]{Neal:2011,Betancourt:2018}, and the latter enables simple and efficient chain-level parallelism. 

In this chapter, we will offer some patterns for accelerating MCMC workflows using these software frameworks running on modern hardware, and show the benefits of doing so. We will also highlight some challenges that can come up.

\section{Background: The classical workflow for probabilistic modeling and Markov chain Monte Carlo}
\label{sec:classical}

Probabilistic inference in statistical models has often followed a similar pattern since at least 2013 \citep{Carpenter:2017}: CPU-based, with a single MCMC run per core, which historically meant running 4--8 chains.
A typical MCMC-based workflow for probabilistic modeling involves many steps \citep{Gelman:2020}, but we focus on two parts: defining a model, and fitting the model.
Defining a model typically involves specifying the model parameters $\theta$, priors over them, $p(\theta)$, as well as the likelihood $p(y \mid \theta)$ where $y$ is a random variable corresponding to observed quantities.

To fit the model, we try to generate samples $\theta_t$ from the posterior $p(\theta \mid y)$, for the purpose of evaluating expectations of some test function $f$, $\mathbb{E}[f(\theta)]$\footnote{All the expectations we consider in this section are conditional on $y$, but we drop this dependence for notational clarity.}, using Monte-Carlo averaging  
\begin{equation} \label{eq:expectation}
    \hat{\theta}_f = \frac{1}{T}\sum_{t=1}^T f(\theta_t).
\end{equation}
Markov Chain Monte Carlo (MCMC) is a class of methods for generating samples which can scale well with the dimension of the space in terms of compute and memory usage. MCMC typically simulates an ergodic Markov chain whose equilibrium distribution is the desired distribution \citep{Bishop:2006}. 

There are many flavors of MCMC, some tailored to specific model classes.
Early software \citep[e.g.,][]{Lunn:2000} focused primarily on Gibbs samplers, but as automatic-differentiation capabilities became more widely available, gradient-based MCMC became more popular \citep{Carpenter:2017,Strumbelj:2023}.
A particularly efficient choice is Hamiltonian Monte Carlo \citep[HMC;][]{duane:1987,Neal:2011,Betancourt:2018,Hoffman:2014}.
HMC only requires access to the log-density (up to a normalizing constant) of the posterior and its gradient, both of which are readily available when using a dedicated library to define the model.
HMC exploits gradient information of the log-density to efficiently generate samples. An implementation is provided in \Cref{sec:hmc}.

\subsection{Measuring efficiency of MCMC}

There are two sources of error in MCMC estimates of expectations: bias and variance. For geometrically ergodic Markov chains, bias becomes exponentially small in the number of initial ``warmup'' samples that are excluded from the final estimate. Variance, on the other hand, shrinks as the reciprocal of the \emph{effective sample size} \citep[ESS;][]{Vehtari:2021}; for the estimator in \Cref{eq:expectation}, the variance is
\begin{equation*}
\operatorname{Var}_{\mathrm{MCMC}}[\hat{\theta}_f] = \frac{\operatorname{Var}_{p(\theta\mid y)}[f(\theta)]}{\operatorname{ESS}_f}.
\end{equation*}
Note that ESS depends on the function $f$ whose expectation we want to estimate.
If we run warmup long enough that variance dominates bias, then maximizing ESS / second minimizes the ``time to acceptable error".
This suggests adapting hyperparameters to maximize ESS, but ESS is difficult to estimate directly from a small number of samples.
Instead, a related quantity, Expected Squared Jump Distance \citep[ESJD;][]{Pasarica:2010}, is sometimes used:
\begin{equation*}
    \operatorname{ESJD}_f = \mathbb{E}\left[ \| f(\theta') - f(\theta)\|^2\right].
\end{equation*}
ESJD only uses only the current state $\theta'$ and previous state $\theta$, so is easy to compute during hyperparameter adaptation.
When $f$ is the identity function $\operatorname{Id}(x) = x$, maximizing $\operatorname{ESJD}_{\operatorname{Id}}$ minimizes the first-order autocorrelation of the MCMC chain, which is closely related to ESS.
The popular No-U-Turn Sampler \citep[NUTS;][]{Hoffman:2014} uses this principle to construct its proposal.

\subsection{Automatic differentiation for gradients of the density}

\label{sec:autodiff}
All modern deep-learning-oriented frameworks provide a higher-order function \lstinline{grad} that transforms a function $f: \mathbb{R}^d \rightarrow \mathbb{R}$ to its gradient $\nabla f: \mathbb{R}^d \rightarrow \mathbb{R}^d$ \citep{JAX:2018,Paszke:2019} when an appropriate gradient exists. There are situations where a gradient calculation will not make sense, e.g., with discrete variables, but the requirement is somewhat weaker than ``continuously differentiable"; for example, the log-density of the Laplace distribution may differentiated despite being only differentiable almost everywhere.

The transformed function $\nabla{f}$ implements reverse-mode automatic differentiation \citep{Griewank:2008}, which, unlike naïve methods like finite differences, is numerically accurate and only incurs a compute cost directly proportional to the cost of the original function $f$.
This means that, at the very least, an implementation of HMC no longer requires the user to provide an implementation of the gradient of the log-density.

\subsection{Automatic differentiation and changes of variables}

Somewhat more subtly, a robust implementation of HMC will allow for densities whose support is not all of $\mathbb{R}^d$. For example, a scale parameter is typically only supported on the positive reals. If $\pi(\theta)$ is supported on $\Omega$, we may use a bijective, differentiable $T: \mathbb{R}^d \rightarrow \Omega$, and enforce the support constraint by the change-of-variables formula:
\begin{equation}
\label{eq:changeofvariable}
z\triangleq T^{-1}(\theta),\qquad \pi(z) = \pi(\theta) \left| \frac{\partial T}{\partial z} (z) \right|,
\end{equation}
where $|\partial T/\partial z|$, the determinant of the Jacobian of the transformation, is computable automatically \citep{Oryx:2022}, through a curated collection of implementations \citep{Carpenter:2017,Dillon:2017}, or some combination of the two. The benefit of considering $\pi(z)$ instead of $\pi(\theta)$ is that we do not need to require an algorithm that works on supports more complicated than $\mathbb{R}^d$.

Note that the choice of transformation is not unique---both softplus (that is, $\theta \mapsto \log(1 + e^x)$) and $\exp$ are bijective maps from $\mathbb{R} \rightarrow \mathbb{R}^+$. Indeed the choice may be important to inference: the transformation may interact with the density and create challenging geometries. Conversely, a transformation may make a density particularly easy to sample from \citep[e.g.,][]{Hoffman:2019}. 

\section{Running example: Hamiltonian Monte Carlo on a sparse Bayesian logistic-regression model in JAX}

We illustrate the process of using accelerators to do MCMC inference by setting up a simple Bayesian logistic-regression model with a sparsity-inducing prior on the coefficients, and then running HMC on it in JAX.
For pedagogical purposes, we provide the implementations using as much raw JAX as possible.
In real-world scenarios it is often preferable to use a probabilistic-programming library (e.g., NumPyro \citep{Phan:2019}, Pyro \citep{Bingham:2019}, PyMC \citep{Abril:2023}, or TensorFlow Probability \citep{Dillon:2017}), and standalone inference libraries such as FunMC \citep{Sountsov:2021a} and BlackJAX \citep{Cabezas:2024} to define the probabilistic models and perform the inference.

\subsection{Probabilistic model and posterior density}
\label{sec:running_example}

We consider the following model:
\begin{align*}
    \tau &\sim \mathrm{Gamma}(0.5, 0.5) \\
    \lambda_d &\sim \mathrm{Gamma}(0.5, 0.5) \\
    \beta_d &\sim \mathcal{N}(0, 1) \\
    y_n &\sim \mathrm{Bernoulli}(\sigma((\tau \lambda \odot \beta)^T x_n))),
\end{align*}
where $\tau$ is a scalar global coefficient scale, $\lambda$ is a vector of local scales, $\beta$ is the vector of unscaled coefficients, $x$ are the features and $y$ are the labels.
$d$ indexes the feature dimensions, while $n$ indexes the examples dimension.
$\sigma$ is the logistic function and $\odot$ is the Hadamard product.
For HMC, we only need to evaluate the joint log-density pointwise, so let us implement this in JAX as a function that takes in the random variable values and returns the log-density:

\begin{lstlisting}
import tensorflow_probability.substrates.jax as tfp
tfd = tfp.distributions

def joint_log_prob(x, y, tau, lamb, beta):
    lp = tfd.Gamma(0.5, 0.5).log_prob(tau)
    lp += tfd.Gamma(0.5, 0.5).log_prob(lamb).sum()
    lp += tfd.Normal(0., 1.).log_prob(beta).sum()
    logits = x @ (tau * lamb * beta)
    lp += tfd.Bernoulli(logits).log_prob(y).sum()
    return lp
\end{lstlisting}

We use the JAX port of the TensorFlow Probability distributions library \citep{Dillon:2017} to implement the primitive log-density computations as that implementation has high numerical precision even when using single-precision floating-point arithmetic (see \Cref{sec:numerics} for further discussion on floating-point precision concerns).
We evaluate the log densities on vector-valued inputs, and then sum them (as the relevant random variables are conditionally independent).
Efficient implementations (such as TensorFlow Probability's) will perform this computation in parallel.
The computation of the log-density of $\lambda$ exhibits \textit{model parallelism}, as the computation scales with the model size (the number of features in this case).
Similarly, the computation of the log-probability of $y$ exhibits both model parallelism and \textit{data parallelism}, as the computation scales with the data size (the number of training examples in this case). See \Cref{sec:parallel} for more details. 

Before we move on, it is convenient to re-express the model in terms of a single vector of latent parameters that are supported on $\mathbb{R}^d$ (the gamma-distributed scale parameters as written above are only supported on $\mathbb{R}^+$).
This involves some simple array manipulation exposed by JAX, and applying the standard change-of-variables formula.
Taking the logarithm of both sides of \Cref{eq:changeofvariable}, we get
\begin{align}
\log \pi(z) = \log \pi(\theta) + \log \left| \frac{\partial T }{\partial z}(z) \right|
\end{align}
While there are many possible choices of $T$, in this case a log-transform is a good choice for the scale parameters (i.e., $T(z)=e^z$, and $\log|\frac{\partial T}{\partial z}(z)| = z$). 
Here is the complete implementation of the unconstrained joint log-density function:
\begin{lstlisting}
import jax.numpy as jnp

def unconstrained_joint_log_prob(x, y, z):
    ndims = x.shape[-1]
    unc_tau, unc_lamb, beta = jnp.split(z, [1, 1 + ndims])
    unc_tau = unc_tau.reshape([])  # Make unc_tau a scalar
    tau = jnp.exp(unc_tau)
    ldj = unc_tau
    lamb = jnp.exp(unc_lamb)
    ldj += unc_lamb.sum()
    return joint_log_prob(x, y, tau, lamb, beta) + ldj
\end{lstlisting}

Lastly, since we're interested in just the posterior of this model, we can condition it via partially applying the \lstinline{x} and \lstinline{y} arguments with some data.
For this example, we'll use the German Credit dataset \citep{german_credit_data}, $N_\mathrm{observations}=1000, N_\mathrm{features}=24$:
\begin{lstlisting}
from functools import partial

target_log_prob = partial(unconstrained_joint_log_prob, x_data, y_data)
\end{lstlisting}
The resulting \lstinline{target_log_prob} function represents an unnormalized version of the posterior $p(z\mid x, y)$; it is a function only of $z$.

So far, the model code does not look so different from R or plain NumPy code.
Now, we will show capabilities which are more unique to JAX and similar toolkits.
First, since HMC requires the gradient of the log-density, we will use JAX to compute it using automatic differentiation.
JAX exposes this feature as a one-line program transformation:
\begin{lstlisting}
import jax

target_log_prob_and_grad = jax.value_and_grad(target_log_prob)
tlp, tlp_grad = target_log_prob_and_grad(z)
\end{lstlisting}
In the last line, we evaluate \lstinline{target_log_prob_and_grad} to compute both the log-density and its gradient with respect to \lstinline{z}.
Recall from \Cref{sec:autodiff} that the \lstinline{target_log_prob_and_grad} uses automatic differentiation to compute an accurate, efficient gradient, as well as computing the log-density for free, as it is a byproduct of computing the gradient.

\subsection{Hamiltonian Monte Carlo}
\label{sec:hmc}

Now we are ready to move on to using HMC.
Below, we implement a simple version of HMC using JAX. In particular, this implementation does not accept or adapt a mass matrix \citep{Neal:2011, Betancourt:2018} (it is implicitly an identity matrix of the appropriate size), nor does it adapt the step size or number of leapfrog steps.
We use the variable \lstinline{z} for the parameters, and \lstinline{m} for the momentum. 
\begin{lstlisting}
def leapfrog_step(target_log_prob_and_grad, step_size, i, leapfrog_state):
    z, m, tlp, tlp_grad = leapfrog_state
    m += 0.5 * step_size * tlp_grad
    z += step_size * m
    tlp, tlp_grad = target_log_prob_and_grad(z)
    m += 0.5 * step_size * tlp_grad
    return z, m, tlp, tlp_grad

def hmc_step(target_log_prob_and_grad, num_leapfrog_steps, step_size, z, seed):
    m_seed, mh_seed = jax.random.split(seed)
    tlp, tlp_grad = target_log_prob_and_grad(z)
    m = jax.random.normal(m_seed, z.shape)
    energy = 0.5 * jnp.square(m).sum() - tlp
    new_z, new_m, new_tlp, _ = jax.lax.fori_loop(
        0,
        num_leapfrog_steps,
        partial(leapfrog_step, target_log_prob_and_grad, step_size),
        (z, m, tlp, tlp_grad))
    new_energy = 0.5 * jnp.square(new_m).sum() - new_tlp
    log_accept_ratio = energy - new_energy
    is_accepted = jnp.log(jax.random.uniform(mh_seed, [])) < log_accept_ratio
    # select the proposed state if accepted
    z = jnp.where(is_accepted, new_z, z)
    hmc_output = {"z": z,
                  "is_accepted": is_accepted,
                  "log_accept_ratio": log_accept_ratio}
    # hmc_output["z"] has shape [num_dimensions]
    return z, hmc_output

def hmc(target_log_prob_and_grad, num_leapfrog_steps, step_size, num_steps, z,
        seed):
    # create a seed for each step
    seeds = jax.random.split(seed, num_steps)
    # this will repeatedly run hmc_step and accumulate the outputs
    _, hmc_output = jax.lax.scan(
        partial(hmc_step, target_log_prob_and_grad, num_leapfrog_steps, step_size),
        z, seeds)
    # hmc_output["z"] now has shape [num_steps, num_dimensions]
    return hmc_output
\end{lstlisting}
The only unusual part of this implementation is that we eschewed the use of Python control flow (\lstinline{if}, \lstinline{else}, \lstinline{for} etc), and instead used API calls like \lstinline{jnp.where} for conditionals and \lstinline{jax.lax.scan} and \lstinline{jax.lax.fori_loop} for for-loops.
\lstinline{jax.lax.scan} is roughly equivalent to this code:
\begin{lstlisting}
def scan(f, state, xs):
  output = []
  for x in xs:
    state, y = f(state, x)
    output.append(y)
  return state, jnp.stack(output)
\end{lstlisting}
The use of these functional forms of control flow is a JAX-specific limitation, and is required for the compiler to generate efficient low-level code.

We can now pass \lstinline{target_log_prob_and_grad} into our HMC implementation and generate an MCMC chain and simple diagnostics:
\begin{lstlisting}
hmc_output = hmc(target_log_prob_and_grad, num_leapfrog_steps, step_size,
    num_samples, z_init, seed)
\end{lstlisting}
To properly diagnose the success or failure of the HMC sampling process we generally need to run multiple independent chains.
This is an opportunity for \emph{chain parallelism}; we can parallelize the independent computations needed to advance the multiple chains.
One way to do this would be to rewrite the \lstinline{hmc} function to accept multiple \lstinline{z_init} values in a \textit{batch},
but this would make the code more complex; most arrays would get saddled with an additional leading (i.e. first) axis corresponding to the multiple chains simulated in parallel.
JAX provides an alternative: an automatic-vectorization program transformation:
\begin{lstlisting}
vhmc = jax.vmap(
    partial(hmc, target_log_prob_and_grad, num_leapfrog_steps, step_size, num_steps))
# z_inits has shape [num_chains, num_dimensions], seeds has shape [num_chains]
hmc_output = vhmc(z_inits, seeds)
# hmc_output["z"] now has shape [num_chains, num_steps, num_dimensions]
\end{lstlisting}
This transformation, called the \textit{vectorized map}, automatically and efficiently transforms existing single-chain HMC implementation to run multiple HMC chains in parallel.
\textit{jax.vmap}, in this basic usage, requires that the function arguments have a leading chain axis and it will add that same axis to the function outputs.
Note that we make sure that each chain gets its own starting location and its own seed, so that the chains are statistically independent.

\subsection{Experiments}
\label{sec:experiments}
To demonstrate the performance characteristics of a GPU on a model like this, we can run NUTS and HMC on the German Credit dataset from the previous section. We use the HMC and NUTS implementations from BlackJAX \citep{Cabezas:2024}, the data and log-density from Inference Gym \citep{Sountsov:2020}, compute diagnostics with ArviZ \citep{Kumar:2019} and use an Nvidia P100 GPU. We run for 1000 MCMC steps in each case, and use a pre-tuned step size, mass matrix, and number of leapfrog steps. We focus on the estimation of the global scale parameter, $\tau$. We run the same experiments using a 28-core Intel Xeon Platinum 8273CL CPU.

\begin{table}[ht]
\centering
\begin{tabular}{|l|l|l|l||l|l|l|} \hline
Chains & Draws & Device & Method & time (s) & ESS & ESS / s \\ \hline \hline
\multirow{4}{*}{64} & \multirow{4}{*}{64,000} & \multirow{2}{*}{GPU} & HMC & 2.4  & 24,041 & 10,017                       \\
& & & NUTS & 4.3  & 16,005 & 3,731  \\ 
& & \multirow{2}{*}{CPU} & HMC & 27.0 & 24,131 & 894 \\
& & & NUTS & 54.4 & 16,048 & 295 \\ \hline
\end{tabular}
\caption{
\label{table:hmc_vs_nuts}
Comparing NUTS and HMC speeds on the sparse logistic regression example with no adaption, using a pre-determined mass matrix, step size, initialization, and trajectory length (for HMC). Note that, taken together with \Cref{fig:hmc_speed} and \Cref{fig:hmc_speed_cpu}, this table depends on the choice of number of chains: the performance comparison would look better for the GPU with more chains, and worse with fewer chains. 
}
\end{table}
\Cref{table:hmc_vs_nuts} shows the relative wall time and effective sample size of HMC and NUTS. As discussed, the control flow overhead from NUTS (see also \Cref{fig:hmc_vs_nuts_profiling} and discussion), along with subsampling from the trajectory to maintain detailed balance with a dynamic trajectory length, means that NUTS takes longer on both measurements. We emphasize that this benchmark is not realistic: we hand-tuned the hyperparameters for HMC so that it would perform well, and this tuning time is not reflected in the benchmark. However, we can consider the promise of being able to automatically tune HMC and see similar gains. 

\begin{figure}[ht]
\centering
\includegraphics[width=0.9\textwidth]{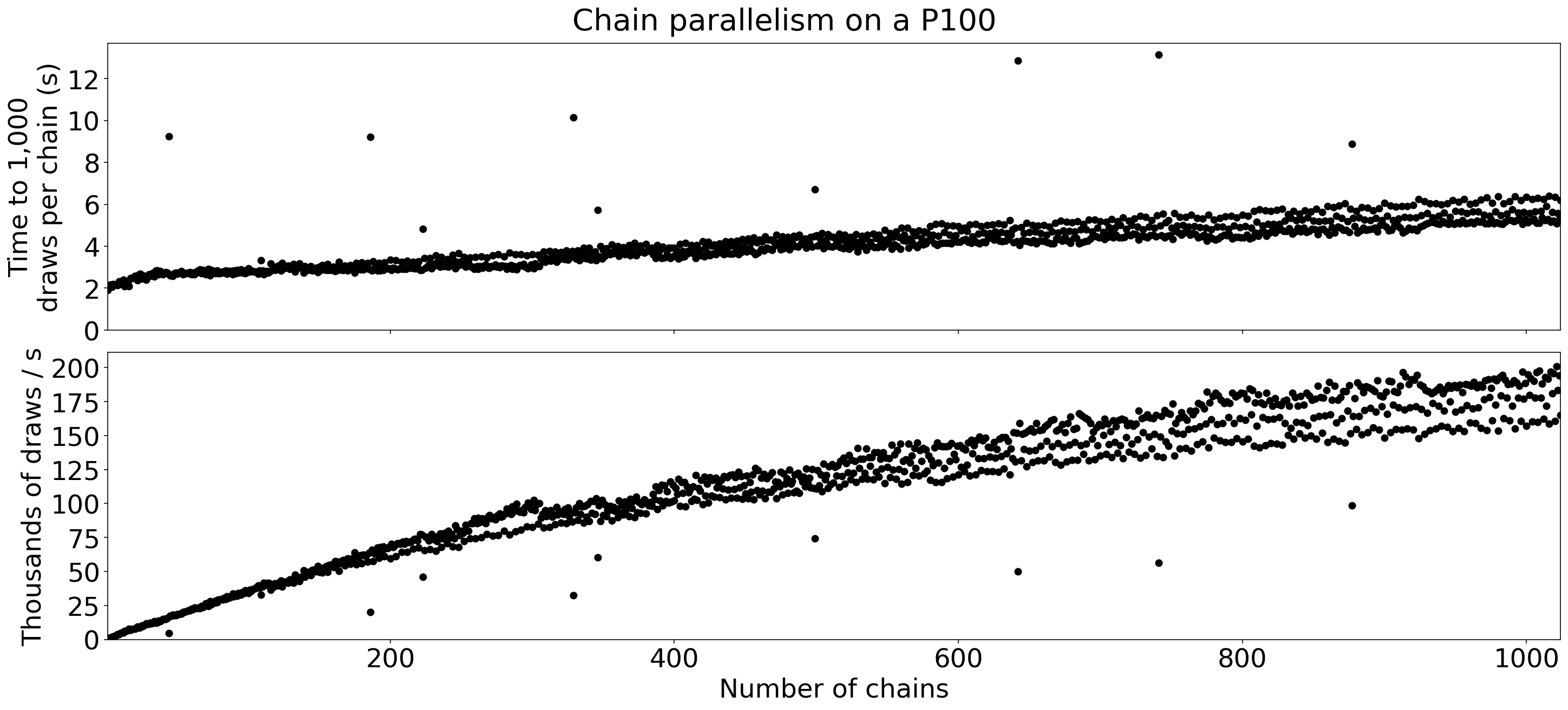}
\caption{\label{fig:hmc_speed}Wall time to run 1000 draws per chain of HMC on a P100 GPU with various numbers of chains, as well as the number of chains per second.}
\end{figure}

\begin{figure}[ht]
\centering
\includegraphics[width=0.9\textwidth]{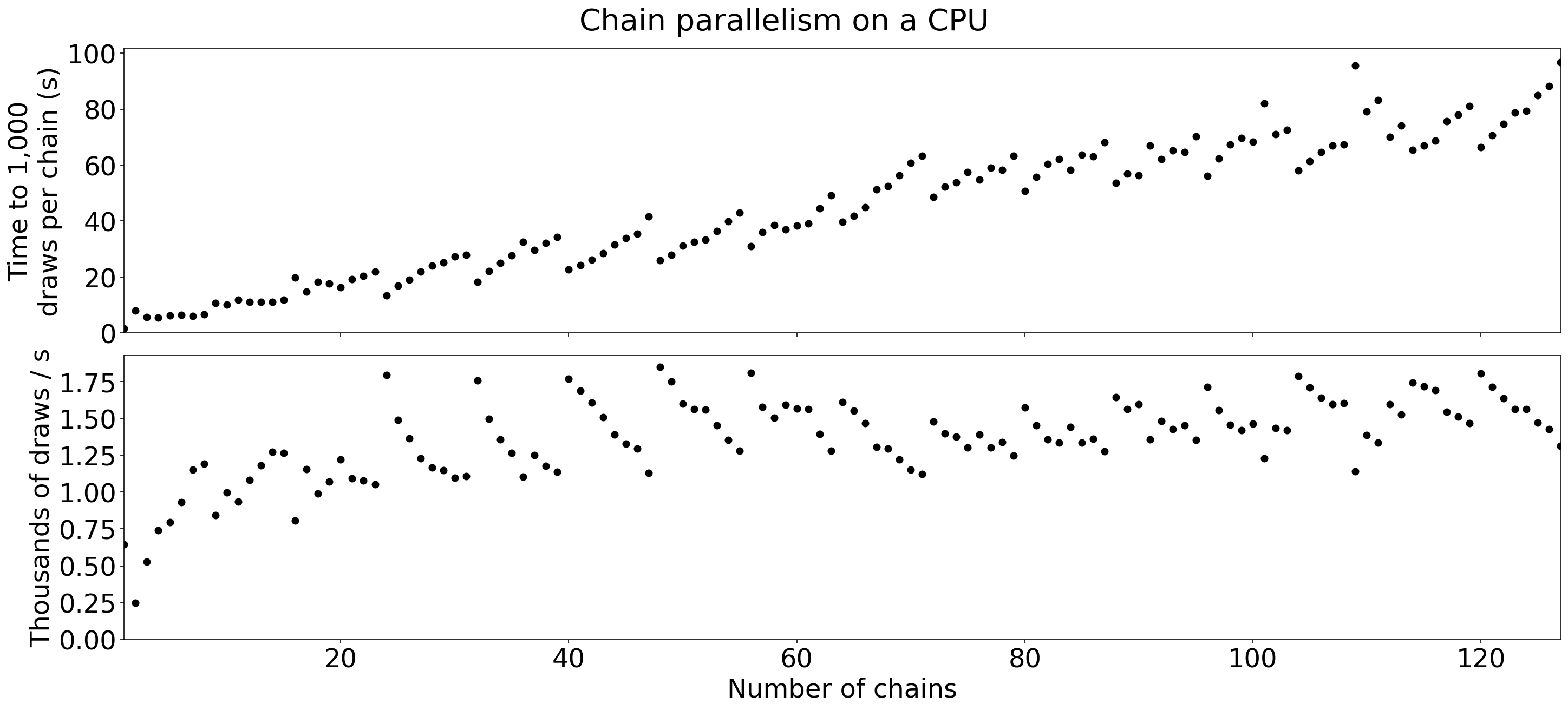}
\caption{\label{fig:hmc_speed_cpu}Wall time to run 1000 draws per chain of HMC on a CPU with various numbers of chains, as well as the number of chains. Note that the draws per second flattens out as all available parallelism becomes exhausted, but does not degrade. }
\end{figure}

We also look at the speed of fitting HMC as we vary the number of chains. Note that the wall time goes up as the number of parallel chains goes up, but this increase is more than offset by the number of draws. In the case of the GPU benchmark in \Cref{fig:hmc_speed}, we in fact see the number of draws per second continuing upwards, as we have failed to exhaust the parallelism offered by the accelerator. The CPU benchmark in \Cref{fig:hmc_speed_cpu} shows the sampling rate levelling off below 2000 draws per second, a small fraction of the number of draws per second on the P100 GPU.

\section{Parallelism opportunities for MCMC workflows}
\label{sec:parallel}

As we saw in the example above, there are (at least) three kinds of parallelization with suitable hardware which come about in a typical MCMC workload:
\begin{itemize}
    \item \textbf{Chain Parallelism:} This may be the most obvious opportunity for parallelism in MCMC workloads. Multiple chains can run in parallel either completely independently, or with (usually minimal) communication to share adaptation signals.
    \item \textbf{Data Parallelism:} We often must compute and sum the log likelihood of many data points, and these computations can often be done in parallel.
    \item \textbf{Model Parallelism:} Our models may include conditional-independence structures that allow for some log densities to be computed in parallel. Also, the state of our Markov chains may be high dimensional, and computations involving different parameters may often be parallelized.
    Model parallelism can be present even when there are no conditional-independence structures in the likelihood.
\end{itemize}
The value of reducing wall-clock time per step by exploiting data- and model-level parallelism is obvious. The value of running many chains in parallel is subtler, but very significant. It enables diagnostics \citep[e.g.,][]{Gelman:1992,Margossian:2023}, offers opportunities for hyperparameter adaptation \citep[e.g.,][]{Gilks:1994,Hoffman:2022}, reduces variance, and reduces bias for some estimands such as quantiles. See \citet{Margossian:2023b} 
for an excellent discussion of the virtues (and challenges) of running many short chains instead of a few long ones.

The Bayesian logistic regression above includes all three kinds of parallelism. We run many chains (chain parallelism); the dataset contains 1000 observations, each of whose likelihoods can be computed in parallel (data parallelism); and the model has dozens of parameters, whose contributions to the likelihood and prior can be evaluated in parallel (model parallelism).

\subsection{How to trade off different kinds of parallelism}

There may be tensions between these kinds of parallelism. For example, we may not be able to implement a computation in parallel along both the data and model axis, or we may have a limited budget of parallel computation, and need to decide how best to allocate it. How to navigate these tradeoffs?

Our answer, at least when working on a single device, is simple: \emph{let the numerical framework decide.}
Frameworks will commonly attempt to take multidimensional arrays and take one or more axes and use multiple threads to perform some computation in parallel.
Thus, the mere act of aggregating values (e.g. the MCMC chain state) into arrays with chain, data and model axes is an avenue to parallelization.
Some frameworks require use of a compilation step (e.g. JAX traces the program in Python to produce a compiled XLA function \citep{Frostig:2018}) to unlock other opportunities, such as running multiple independent computations in parallel.

For example, in the Bayesian logistic regression example above, the bulk of the log-density computation ($N_\mathrm{chains}\times N_\mathrm{dimensions} \times N_\mathrm{observations}$ multiplications and additions) is naturally implemented as a matrix multiplication. The authors of frameworks like JAX and PyTorch work \emph{very} hard to make large matrix multiplications efficient, since these operations dominate the cost of neural-network training and evaluation. There is no need for us to reinvent that wheel.

That said, cloud-computing providers now offer the ability to run computations on multiple devices networked together by very high-speed connections.
To use these multi-device setups, the user must take a more active role in sharding data and computation across devices.
A full discussion of multi-device sharding is beyond the scope of this chapter, but generally frameworks will have tools to simplify the process.
Whenever feasible, we recommend sharding by \emph{chain}, since this is simplest to implement and involves the least cross-device communication.
Frameworks like JAX support sharding along multiple axes at once, which enables sharding along data and model axes as well, in which case sharding on the data axis is common next choice.
Sharding along the model axis is typically reserved for largest of models, such as large Bayesian Neural Nets (BNNs), where a single model might not fit into memory of a single device.

As an example of of multi-device MCMC consider the large study by \cite{Izmailov:2021} used 512 TPUs to perform full-batch HMC inference on a BNN.
In that work the datasets were not excessively large, but the problem required 10,000 to 20,000 leapfrog iterations.
To manage the compute cost, the data axis was sharded across devices.
The chain axis was not used (multiple chains were run sequentially) and the neural net was not large enough to warrant sharding along the model axis.

In addition to exact MCMC, there are also ``divide-and-conquer'' style methods which perform MCMC on shards of the whole dataset and then combine the resultant posteriors, typically by first estimating the posterior density for each shard and then multiplying them (see \cite{Bardenet:2017} for a review).
Such methods also benefit from framework support, while also enjoying a reduced cross-device-communication cost compared to exact MCMC in exchange for some bias.

\subsection{Advances in many-chain MCMC}
\label{sec:recent_work}

Access to a large number of MCMC chains enables new convergence diagnostics and hyperparameter adaptation techniques. We refer to the work by \cite{Margossian:2023b} 
for a helpful discussion on this, and provide some highlights here.

There have been recent advances in making HMC more efficient on parallel accelerators. Recall that in order to run HMC, the mass matrix, step size, and trajectory length must be chosen as hyperparameters. 

For step-size and mass-matrix adaptation, it is straightforward to extend the commonly used stochastic-optimization algorithms by computing the relevant empirical averages across chains in addition to across time.
This is not without caveats, however.
For step size, it is common for an HMC chain to get ``stuck'', meaning that nearly all proposals are rejected due to difficult local geometry of the log-density, and a small step size is required for the chain to escape that bad region.
If an arithmetic mean is used to estimate the current acceptance rate, and only a few chains are stuck, the step size will not get small enough.
Instead of the arithmetic mean, \cite{Hoffman:2021} propose using the harmonic mean to compute the empirical cross-chain acceptance probability.
The harmonic mean of a set of positive values will be far closer to its smallest member, causing the ensemble of chains to ``slow down'' for the stragglers.

For trajectory length, NUTS has implementation and hardware drawbacks discussed in \cref{sec:simd} and \Cref{sec:experiments}, which motivates a search for alternatives.
The many-chain regime enables the use of gradient-based estimators that target the gradient of $\operatorname{ESJD}_{\operatorname{Id}}$ with respect to the trajectory length.
Algorithms maximizing $\operatorname{ESJD}_{\operatorname{Id}}$ 
are good at estimating means, but not necessarily at estimating variances. 
The Change in the Estimator of the Expected Square \citep[ChEES;][]{Hoffman:2020b}, is invariant to shifts and rotations, relatively insensitive to directions of low posterior variance (since the work done in high-variance directions will address those), and focused on estimating the variance: 
$$
\operatorname{ChEES} =  \frac{1}{4} \mathbb{E}\left[ \left( \| \theta' - \mathbb{E}[\theta]\|)^2 - \|\theta - \mathbb{E}[\theta]\|^2\right)^2\right] = \frac{1}{4}\operatorname{ESJD}_{\|z\|^2},
$$
where $z := \theta - \mathbb{E}[\theta]$.
Using more chains allows expectations to be calculated across chains during an adaptation phase, and ChEES takes a gradient through this objective function with respect to the trajectory length in order to optimize it. 
Further improvements to the metric improved its robustness and efficiency \citep{Sountsov:2021,Riou-Durand:2023}, all relying on multi-chain estimators.

\begin{figure}[ht]
  \begin{center}
  \includegraphics[width=4in]{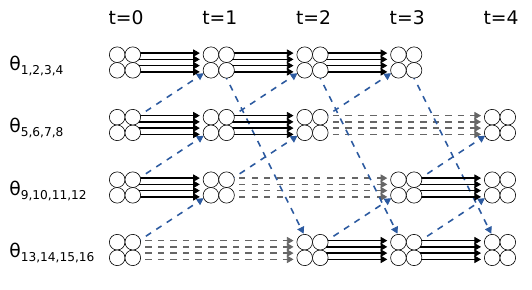}
  \end{center}
  \caption{
    Graphical model illustrating $K$-fold ECA.
    The states are split into $K$ folds of $N$ states each,
    and each fold $k$ is updated using parameters computed
    from its neighbor fold $k + 1 \mod K$. Each iteration
    we skip the update for a different fold. Solid black
    lines denote MCMC updates, dashed blue lines denote
    dependence through kernel parameters, dashed gray
    lines denote skipping an update.
  }
  \label{fig:eca}
\end{figure}

When the amount of chain parallelism available is very large it is possible to avoid stochastic optimization altogether. \cite{Hoffman:2022} introduce an ensemble-chain adaptation (ECA) scheme, which allows updating one chain using statistics that depend on other chains.
This can be done without violating detailed balance, unlike the adaptation schemes described previously which change the stationary distribution \citep{Andrieu:2008}.
ECA is also amenable to a K-fold version, which may further take advantage of hardware batching, illustrated in \Cref{fig:eca}. 

It is important to verify that the MCMC chain has converged to its stationary distribution before using its generated samples for downstream analysis.
$\hat{R}$ is a popular family of diagnostics for this purpose which compares within- and across-chain variances \citep{Gelman:1992, Vehtari:2021}.
While the across-chain variance computation works well in the many-chain regime, computation of within-chain variance is an inherently sequential task.
\cite{Margossian:2023} propose nested-$\hat{R}$ which carefully partitions chains into superchains, and then compares within- and across-superchain variance, both of which can be computed in parallel.

\section{Challenges}

MCMC workflows can map very well to parallel accelerators, but
there are some pitfalls unique to these devices that have to be kept in mind to fully exploit their performance.
Not every MCMC application will run into these issues, and some of these issues have ready-made solutions (see \Cref{sec:recent_work}) or affect only researchers who wish to develop new methods.
That said, this section provides a mental model for these issues, some ways to realize they are present, and recommendations for how to address them in case relying on the numerical frameworks' built-in parallelization behavior is not enough.

\subsection{Numerics}
\label{sec:numerics}

The accelerators we have considered in this chapter (GPUs, TPUs) do not prioritize double-precision floating point performance, and are more efficient when using single- and half-precision (and below) numerics.
This can have catastrophic interactions with naïve MCMC implementations that take numerical precision for granted.

We already encountered this when we were implementing the example regression problem \Cref{sec:running_example}, where we used log-space computation to preserve precision.
For example, many frameworks parameterize the Bernoulli distribution via the probability parameter.
This can easily cause issues in single-precision due to underflow for probabilities near zero, so it is preferable to parameterize it via the logit-transformed probability.

Another common issue is roundoff error when computing large sums of log densities of random variables to compute the joint log-density.
The roundoff error interacts poorly with MCMC that uses a Metropolis-Hastings accept-reject step, as that requires comparing the log-density evaluated at two points \citep{Hoffman:2020b}.
The single-precision floating point format uses 24 bits to represent the mantissa, which means that when the absolute value of the log-density exceeds $2^{24} = 16777216$, the absolute errors start to become $\ge 1$.
This causes problems with Metropolis-Hastings accept-reject step, as errors on that order are appreciable.
Worse, if the roundoff error causes the MCMC chain to find a state where the error makes the log-density too high, MCMC will have trouble moving away from that state, unless it finds, by chance, a state where the error is similarly high.
States where the roundoff error is low will be rejected, even if they would have been accepted if double precision were used.
This not only introduces bias to the MCMC results, but also makes the chain struggle to make progress.

The first step in addressing numerics problems is verifying that this is the problem in the first place.
Low MCMC acceptance rates can be caused by a multitude of problems unrelated to numerics, e.g. poorly set hyperparameters, poor conditioning of the log-density function, bad choice of parameterization, etc.
The simplest method is to attempt to run the MCMC inference in double-precision.
On GPUs, double-precision comes at a performance and memory use penalty, but oftentimes this can be a good tradeoff if it identifies the issue.
In general, if the MCMC diagnostics look fine after switching to double-precision, then precision-related numerics are a likely culprit.

Roundoff errors in Metropolis-Hastings accept-reject steps can be detected by either looking at the absolute magnitude of the log-densities involved, or by checking for signs of low precision in the logarithms of the accept ratios (e.g., \lstinline{log_accept_ratio} in \Cref{sec:hmc}); for example, log-ratios ending in exactly $.0$, $.5$, or $.25$ should arouse suspicion.
One solution to this problem is to rewrite the computation of the log-acceptance probability to perform a sum of differences of the log densities.
For example, for the likelihood term, we can do the following rewrite:
\begin{align}
    \sum_{i=1}^{N} \log p(y_i \mid \theta') - \sum_{i=1}^{N} \log p(y_i \mid \theta) = \sum_{i=1}^{N} \log p(y_i \mid \theta') - \log p(y_i \mid \theta).
\end{align}
The Python implementation is straightforward, if a bit tedious to write.
Here is a snippet of the function to compute a numerically stable log-density ratio:
\begin{lstlisting}
def joint_log_prob_ratio(x, y, tau_1, tau_2, lamb_1, lamb_2, beta_1, beta_2):
    tau_dist = tfd.Gamma(0.5, 0.5)
    lpr = tau_dist.log_prob(tau_1) - tau_dist.log_prob(tau_2)
    lamb_dist = tfd.Gamma(0.5, 0.5)
    lpr += (lamb_dist.log_prob(lamb_1) - lamb_dist.log_prob(lamb_2)).sum()
    beta_dist = tfd.Normal(0., 1.)
    lpr += (beta_dist.log_prob(beta_1) - beta_dist.log_prob(beta_2)).sum()
    logits_1 = x @ (tau_1 * lamb_1 * beta_1)
    logits_2 = x @ (tau_2 * lamb_2 * beta_2)
    lpr += (tfd.Bernoulli(logits_1).log_prob(y) - tfd.Bernoulli(logits_2)).sum()
    return lpr
\end{lstlisting}
Using a fully-featured probabilistic programming language can simplify the construction of such a function.
After repeating the unconstraining and conditioning steps we performed in \Cref{sec:running_example} for this function as well, we can use it inside the HMC implementation in \Cref{sec:hmc} to compute the log-density ratio:
\begin{lstlisting}[firstnumber=19]
log_accept_prob = (0.5 * (jnp.square(m) - jnp.square(new_m)).sum()
                   + target_log_prob_ratio(new_z, z))
\end{lstlisting}

Other precision-related problems tend to be harder to discover.
A reasonable strategy is to run small parts of the model in higher precision to isolate the problem areas: it is common that only a small part of a model needs to be run at higher precision for the end-to-end inference procedure to succeed.
After the problem area is identified, it is usually possible to rewrite it in a way that fully utilizes single-precision computation by utilizing log-space computation, analytic simplification, and other techniques from numerical analysis.

\subsection{SIMD}
\label{sec:simd}

Hardware accelerators derive much of their performance from using ``wide'' Single-Instruction-Multiple-Data (SIMD) instructions which operate on arrays of floats in parallel.
The ``Multiple'' in SIMD implies a certain minimum level of parallel computation that efficiently uses the hardware.
On CPUs, it is typical for SIMD instructions to handle 4 to 8 floats in parallel.
Code utilizing such SIMD instructions is relatively difficult to write, so in practice SIMD is relegated to highly efficient implementations of primitive matrix-vector operations.
The main parallelism avenue on CPUs that is accessible to practitioners is via multiple threads, each of which can run separate instances of computation in parallel.
For floating-point-heavy code it is typical to place one thread per core, and CPUs typically have anywhere from 4 to 64 cores.

By contrast, GPUs use a type of SIMD execution model called the Single-Instruction-Multiple-Threads (SIMT) which groups computation into blocks (called ``warps'' or ``wavefronts'' depending on the manufacturer) which execute 32 or 64 threads in parallel \citep{Lindholm:2008} and lockstep.
Modern GPUs support multiple warps and can execute thousands of threads in parallel.
Additionally, writing code under the SIMT model is typically simpler than SIMD or multi-threaded CPU due to the lockstep execution.

TPUs do not use SIMT, but instead use very wide SIMD matrix-matrix multiply instructions \citep{Jouppi:2017, Jouppi:2023}.
As a result, TPUs are best used for matrix-multiplication-heavy models such as hierarchical linear models or Bayesian neural networks.

Accelerators typically run at slower clock speeds than a contemporaneous CPU which makes them prefer massively parallelizable code with few sequential operations.
In this regime, accelerators can perform orders of magnitude more FLOPs than running on a CPU.

When running many-chain MCMC, it often makes sense to place individual chains onto the individual threads.
On a GPU this means means that running 32 to 64 threads can be effectively ``free'' as fewer threads would not fill up the minimum computation size.
However, due to the lockstep execution requirement, this only works well when separate MCMC chains perform the same computation at the same time across all threads \citep{Lao:2020}.
This can be violated for algorithms which have control flow or are adaptive in a way that makes the chains differ in the amount of computation they perform.
This phenomenon is called ``warp divergence''.

For example, consider a pair of chains that use HMC, one of which intends to take one leapfrog step, and the other which intends to take two.
The SIMD execution model would parallelize this by running both chains for two leapfrog steps, but then discarding the result of the second leapfrog step from the chain that intended to take only one.
It is clear that this approach wastes computation, and sometimes can turn a parallel algorithm into a sequential one.

To take advantage of parallel hardware efficiently we need to adapt algorithms to respect these restrictions, and sometimes we need to choose a different algorithm altogether.
For example, in \Cref{sec:running_example} we implemented a simple version of HMC.
Amongst many issues with that implementation, a glaring one is that it takes the same number of leapfrog steps in each HMC iteration.
This is known to cause problems due to resonance of the Hamiltonian dynamics \citep{Neal:1996,Hoffman:2021}.
A simple solution to this that does not affect the integration accuracy (and thus the acceptance probability of the proposal) is to jitter the number of leapfrog steps taken in each HMC step.
A straightforward way to implement this, would be to alter the following lines (changes highlighted in red):
\begin{lstlisting}[firstnumber=30]
def hmc_step(target_log_prob_and_grad, num_leapfrog_steps, step_size, z, seed):
    m_seed, ~jitter_seed~, mh_seed = jax.random.split(seed~, 3~)
    ~num_leapfrog_steps = jax.random.randint(~
        ~jitter_seed, [], minval=1, maxval=1 + 2 * num_leapfrog_steps)~
    tlp, tlp_grad = target_log_prob_and_grad(z)
\end{lstlisting}
When transformed via \lstinline{jax.vmap}, however, this produces warp divergence when using chain parallelism.
Different chains may choose a different number of leapfrog steps, while the computation will always run for the largest number of leapfrog steps chosen by any chain.

The solution to this issue is to make every chain use the same random seed for the jittered value.
In terms of implementation, one option is to add another seed argument that we deliberately make equal across chains.
An example implementation might be:
\begin{lstlisting}[firstnumber=30]
def hmc_step(target_log_prob_and_grad, num_leapfrog_steps, step_size, z,
        ~seed_and_jitter_seed~):
    ~seed, jitter_seed = seed_and_jitter_seed~
    m_seed, mh_seed = jax.random.split(seed)
    num_leapfrog_steps = jax.random.randint(
        ~jitter_seed~, [], minval=1, maxval=1 + 2 * num_leapfrog_steps)
    tlp, tlp_grad = target_log_prob_and_grad(z)
    ...
    
def hmc(target_log_prob_and_grad, num_leapfrog_steps, step_size, num_steps, z,
        seed~, jitter_seed~):
    seeds = jax.random.split(seed, num_steps)
    ~jitter_seeds = jax.random.split(jitter_seed, num_steps)~
    _, hmc_output = jax.lax.scan(
        partial(hmc_step, target_log_prob_and_grad, num_leapfrog_steps, step_size),
        z, ~(seeds, jitter_seeds)~)
    return hmc_output

vhmc = jax.vmap(
    partial(hmc, target_log_prob_and_grad, num_leapfrog_steps, step_size,
        num_steps),
    ~in_axes=(0, 0, None)~)  # None means that each chain gets the same jitter_seed
hmc_output = vhmc(z_inits, seeds~, jitter_seed~)
\end{lstlisting}

\begin{figure}[ht]
  \begin{center}
  \includegraphics[width=5.5in]{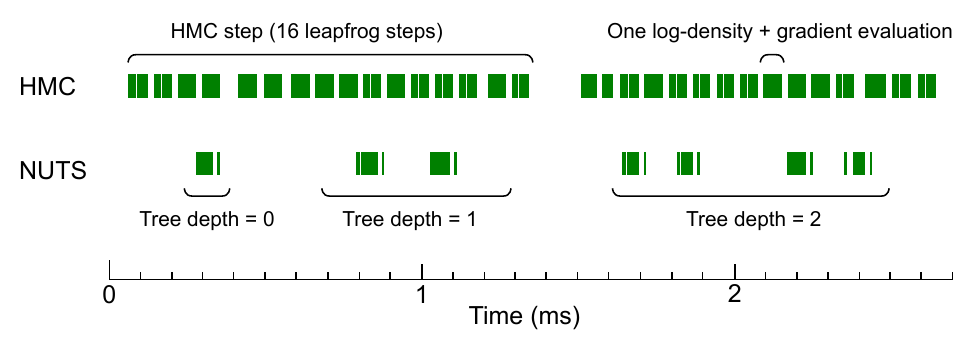}
  \end{center}
  \caption{
    Profiling traces of HMC and NUTS when sampling from the sparse logistic regression model.
    The green rectangles indicate the time spent evaluating the log-density and its gradient.
    The remaining time is taken up by the leapfrog integration, storage of the samples in the output buffer, and, in case of NUTS managing the internal tree proposal stack memory and making tree doubling decisions.
    Overall, NUTS spends a lot of time managing its internal state, reducing the number of leapfrog steps it can do per second compared to HMC.
  }
  \label{fig:hmc_vs_nuts_profiling}
\end{figure}
Sometimes the changes necessary to make an algorithm SIMD-friendly are more involved.
As discussed earlier, NUTS is a variant of HMC that automatically chooses how many leapfrog steps to take based on the local log-density geometry \citep{Hoffman:2014}.
This procedure makes NUTS very robust to different probabilistic models, as it can adapt to the local scale of the log-density.
Even for well-behaved distributions, NUTS saves the practitioner from having to tune the number of leapfrog steps.
All this has caused a lot of effort to be spent making NUTS work on parallel hardware \citep{Lao:2020,Phan:2019,Radul:2020}.
Warp divergence is a given with NUTS by the nature of the algorithm: different chains typically use proposals with different numbers of leapfrog steps, meaning that chains that use fewer than the maximum number will waste some computation evaluating the leapfrog steps up to that maximum and then discarding them.
Even discounting this issue, NUTS requires a lot of control flow and maintenance of a stack to generate its proposal.
The control flow and the additional memory usage, as well as memory copies to read/write from the stack are slow in simple implementations and are complex to optimize.
In many cases, it has been found that a well-tuned HMC, akin to what we implemented in \Cref{sec:hmc}, will outperform NUTS in wall-clock time due to these issues \citep[\Cref{table:hmc_vs_nuts}, and][]{Hoffman:2021}.
For example, we profiled a JAX implementation of NUTS and HMC in \Cref{fig:hmc_vs_nuts_profiling}.
NUTS has to spend an appreciable amount of time managing its state.
HMC does not have as much state management, and can spend a majority of its time evaluating the log-density function and its gradients.
In general, sometimes it is not possible to fully adapt an algorithm to be efficient on parallel hardware, and a different algorithm is required.

\subsection{Memory}
\label{sec:memory}

An important component of maximizing the performance of accelerators is to avoid transferring data between the accelerator's memory and the main CPU memory.
In particular, a common technique for managing the memory requirements of storing MCMC iterates is to stream them to disk.
While keeping things in memory is much faster than offloading to disk, this typically is not a huge problem thanks to considerable operating system support to make this happen reasonably asynchronously (via careful software and hardware caching, memory-mapped files, and other related techniques).
This becomes less favorable when using accelerators, and often deep-learning frameworks do not automate this process
even if the relevant hardware support exists (e.g. via RDMA).
As a result, it becomes more attractive to attempt to perform the analysis, such as computing statistics about the posterior distribution, wholly on the accelerator.

Unfortunately, parallel accelerators typically have less memory (termed VRAM in the context of GPUs) than a CPU, especially a CPU on a workstation computer, meaning that naïvely storing all the MCMC iterates may fail.
Fortunately, developments in streaming statistics (typically coming from online and database contexts) can be used to relieve these memory pressures.
For example, Welford's scheme \citep{Welford:1962} can be used to compute first and second moments, which conveniently extends to the downstream convergence diagnostics such as $\hat{R}$.
Even more advanced variants such as split-$\hat{R}$ are amenable to this by selectively choosing which segments of the MCMC chain to incorporate into the variance estimates.
Quantiles can be estimated using approximate techniques from \cite{Chen:2020}, some of which are amenable to GPUs \citep{Govindaraju:2005}.

\section{A recipe for effective utilization of parallel hardware and frameworks}

In this chapter we discussed how HMC, a popular MCMC method, maps naturally to modern parallel hardware and deep-learning libraries thanks to the latter's support for automatic vectorization and parallelization, as well as automatic differentiation. We presented the following general recipe for the effective use of MCMC on parallel hardware:
\begin{description}
    \item[Use a framework that helps. (\Cref{sec:classical})] Specifically, one that integrates with automatic differentiation and accelerators. These frameworks help by facilitating or getting out of the way of chain/data/model-parallelism, and allow interoperability with a broader ecosystem.
    \item[Given a choice, prefer HMC. (\Cref{sec:classical})] While there are certainly situations where a custom sampler, or even random-walk Metropolis, might be the most efficient sampler, the easy access to automatic differentiation makes it easy to try a gradient-based sampler right from the start. Many pre-built inference libraries also include the latest advances, either in new algorithms, adaptation and convergence diagnostics.
    \item[Let the framework parallelize the computation (\Cref{sec:parallel})] Modern software frameworks like JAX and PyTorch are very good at figuring out how to efficiently run numerical computations on a single GPU or TPU. Vectorized-map transformations can automatically transform single-chain MCMC implementations into efficiently parallelized multi-chain MCMC implementations.
    \item[Be mindful of numerical precision. (\Cref{sec:numerics})] Since single-or-less-precision is the norm on GPUs and TPUs, it is especially important to work in log space---which a reasonable framework will do by default. If the magnitude of the target log-density gets very large, watch out for roundoff error. Enable double-precision for debugging. 
    \item[Be mindful of SIMD. (\Cref{sec:simd})] Prefer inference algorithms with little control flow (like non-NUTS variants of HMC). For models with control flow, consider alternate parameterizations that are more amenable to the hardware constraints. 
    \item[Be mindful of memory use. (\Cref{sec:memory})] When scaling to hundreds of chains, memory use may also be scaled hundreds of times. Streaming algorithms (for example, Welford when computing mass matrices) may be used to reduce the footprint of a program.
\end{description}

Even without using a dedicated probabilistic modeling and inference software library, deep-learning libraries provide enough tools to quickly get started and make efficient use of the available hardware.
The greatest benefit is obtained, however, when taking advantage of software written with the parallel hardware in mind, as it will often surface recent algorithmic and methodological research.
While additional research will no doubt improve the ease-of-use, robustness and efficiency of MCMC algorithms running on parallel hardware, the state of support for such a configuration is already quite good: if a practitioner has access to such hardware and a problem that can be solved by HMC, there are no further real roadblocks to the practical and efficient exploitation of parallel computation.

\section{Acknowledgments}

We thank Bob Carpenter, Charles C. Margossian, and Du Phan for helpful comments.

\bibliographystyle{apalike} 
\bibliography{handbook}

\end{document}